\documentclass[aps,prd,twocolumn,groupedaddress]{revtex4}
\usepackage{amsfonts}

\def\boldsymbol#1{\setbox0=\hbox{$#1$}%
 \kern-.025em\copy0\kern-\wd0
 \kern.05em\copy0\kern-\wd0
 \kern-.025em\raise.0433em\box0 }

\begin{document}

%\preprint{}

\title{Singularity-Free Electrodynamics for Point  Charges and  Dipoles:\break  Classical
Model for Electron Self-Energy and Spin}


\author{S. M. Blinder}
\email[]{sblinder@umich.edu}
%\thanks{}
\affiliation{University of Michigan \break Ann Arbor, MI 48109-1055 USA}


\date{\today}

\begin{abstract}
\noindent It is shown how point
charges and point dipoles with finite self-energies can be accomodated into classical
electrodynamics.  The key idea is the introduction of
constitutive relations for the electromagnetic vacuum, which actually mirrors the
physical reality of vacuum polarization.  Our results reduce to conventional
electrodynamics for scales large compared to the classical electron radius $r_0\approx
2.8\times10^{-13}$ cm.  A classical simulation for a structureless electron is proposed,
with the appropriate values of mass, spin and magnetic moment.
\end{abstract}

%\keywords{}

\maketitle

\section{Introduction}

The most elementary problems in classical electrodynamics are likely to involve point
charges, point dipoles and other tractable mathematical idealizations of physical
reality.  But the simplicity of these solutions often obscures a subtle but pervasive
conceptual flaw.  This appears even in the most elementary formula of electrostatics,
Coulomb's law of force between two point charges $q_1$ and $q_2$ separated by a
distance $r$
\begin{equation}F={{q_1\,q_2}\over r^2}\end{equation} with the corresponding
energy of interaction
\begin{equation}W={{q_1\,q_2}\over r} \end{equation} What if we want to know the
hypothetical force a point charge would exert on {\it itself}?  Since we would then have
$r=0$ in these equations, this force becomes infinitely large, as does the {\it
self-energy}---the energy of  interaction of the point charge with its own electric field.  For
the most part, these difficulties have been ``swept under the rug" since
electrically-charged bodies in real life have a finite size.  A point charge is perhaps just
an abstraction.  Still, it would be highly desirable, even for purely aesthetic reasons, to
remove this imperfection from the otherwise beautiful and complete edifice of
Maxwell's electrodynamics.  Dimensionlesss point charges are in fact the paradigm
for representing fundamental particles in quantum mechanics and quantum field theory. 
And all experimental evidence from high-energy physics appears to support such models.
For example, recent high-energy electron-positron scattering experiments imply an upper
limit of $2\times 10^{-16}\,{\rm cm}$ on the radius of the electron.

The strategy we pursue invokes {\it constitutive relations}.  Usually, these are
phenomenological parameters which represent properties of matter, serving as {\it
inputs} to Maxwell's equations, not implied by the stucture of electrodynamics itself.   In
certain favorable instances, these parameters can be determined theoretically from
quantum theories of matter. The idea which we will exploit is to attribute constitutive
properties to the {\it vacuum}. According to quantum field theory, the ultramicroscopic
vicinity of an elementary charged particle is a seething maelstrom of virtual
electron-positron pairs (and other particles and antiparticles) flashing in and out of
existence.  To take account of this well-established physical reality, a phenomenological
representation for vacuum polarization is introduced within the framework of classical
electrodynamics. As we will show, such a model enables a consistent picture of classical
point charges with finite electromagnetic self-energy.  We must emphasize that the model is
intended in a purely classical context and will not necessarily be in agreement with details
of quantum electrodynamics. In the same sense, continuum models for dielectric media can
be extremely successful without taking account of the underlying atomic nature of matter.

\section{Self-Energy of a Point Charge}

 The energy contained in an electromagnetic field is given by
\begin{equation}W={1\over{8\pi}}\,\int\,\left({\bf E\cdot D + B\cdot H}\right)\,d^3{\bf r}
\end{equation}
In a rest frame, the field produced by a point charge $e$ in vacuum is represented by
${\bf D}={\bf E}=e{\bf\hat r}/r^2$,  ${\bf B}={\bf H}=0$. For concreteness, let 
the particle be an electron. The electromagnetic self-energy is then given by
\begin{equation}W= {1\over{8\pi}}\,\int\,{e^2\over{r^4}}\,4\pi r^2\, dr = \infty
\end{equation} 
The result is infinite unless a lower cutoff is introduced---in which case the electron
acquires a finite size, as in the models proposed by Thomson, Lorentz,
Abraham and others a century ago[1].  With a
radius of the order of $r_0=e^2/mc^2\approx 2.818\times 10^{-13}$ cm, known as
the classical or Thomson radius, the electromagnetic self-energy can
be adjusted to equal $mc^2$. This is in accord with the original idea of Lorentz and
Abraham that the electron's rest mass is purely electromagnetic in origin.  Because of
mutual repulsions among the electron's elements of charge, the whole structure might be
expected to blow itself apart. Poincar\'e postulated the existence of nonelectromagnetic
attractive forces---later called {\it Poincar\'e stresses}---to somehow counterbalance the
Coulomb repulsion.

It was suggested in the 1930's by Furry and Oppenheimer[2] that
quantum-electrodynamic effects could give the vacuum some characteristics of a
polarizable medium, which Weisskopf[3]  represented phenomenologically by an
inhomogeneous electric permittivity, viz.,
\begin{equation}{\bf D}(r) = \epsilon(r){\bf E}(r)\end{equation} 
Thus, assuming the electron's rest mass is entirely electromagnetic,
\begin{equation}
W={1\over{8\pi}}\,\int_0^\infty\,{1\over\epsilon(r)}\,{e^2\over{r^4}}\,4\pi
r^2\, dr=mc^2
\end{equation}
The net charge density $\rho(r)$, taking account of the conjectured
vacuum polarization, is given by
\begin{equation}
\nabla\cdot{\bf E}=-\frac{e
\epsilon'(r)}{ r^2[\epsilon(r)]^2} =4\pi\rho(r)
\end{equation}
The original point charge is here exactly cancelled by a deltafunction contribution
from the polarization charge.  A functional form for $\epsilon(r)$ can be determined
if  the charge density $\rho(r)$ is assumed to be proportional to
the electromagnetic energy density, so that
\begin{equation} -\frac{
\epsilon'(r)}{4\pi r^2[\epsilon(r)]^2}=\frac{e^2}{8\pi m c^2\epsilon(r)r^4}
\end{equation}
The result is[4]
\begin{equation} \epsilon(r)=\exp\left(\frac{e^2}{2mc^2 r}\right)=
\exp\left(\frac{r_0}{2\, r}\right)\end{equation}
The self-energy then follows from
\begin{equation}W=\frac{e^2}{2}\,\int_0^\infty\,\frac{e^{-r_0/2r}}{r^2}\, dr=
\frac{e^2}{r_0}=mc^2
\end{equation}

\section{Self-Energy of a Dipole}
A point electric dipole $\boldsymbol{\mathfrak p}$ located at the origin and directed
along the polar axis produces an axially symmetric field given by
\begin{equation} {\bf D}=\frac{2\, {\mathfrak p} \cos\theta}{r^3}\,{\bf \hat r}+
\frac{ {\mathfrak p} \sin\theta}{r^3}\,\boldsymbol {\hat\theta}
\end{equation}
If it is assumed that the permittivity $\epsilon(r)$ remains spherically symmetrical,
the field energy integrated over solid angle is given by
\begin{equation}\frac1{8\pi}
\int_0^{2\pi}\int_0^{\pi}\,{\bf E\cdot D}\,\sin\theta\,d\theta\,d\phi=
\frac{2\,{\mathfrak p}^2}{ r^6 \epsilon(r)}
\end{equation}
The charge density analogous to (7) is determined by the proportionality 
\begin{equation}\rho(r)=\frac{1}{4\pi}\nabla\cdot{\bf E}\approx
-\frac{\epsilon'(r)}{r^3 [\epsilon(r)]^2}
\end{equation} so that the relation analogous to (8) implies a permittivity of the
form
\begin{equation} \epsilon_{\scriptscriptstyle\rm dipole}(r)=\exp(k^2/r^2)
\end{equation}
where $k$ is a parameter with dimensions of length.  For example,  a hypothetical
electric dipole
$\boldsymbol{\mathfrak p}$ with electromagnetic self-energy $M c^2$ would imply
\begin{displaymath}k=\left(\frac{{\mathfrak p}^2\sqrt\pi}{4 M c^2}\right)^{1/3}
\end{displaymath}

The treatment of a magnetic dipole $\boldsymbol{\mathfrak m}$ is closely
analogous. The magnetic field in vacuum is give by
\begin{equation}
{\bf H}=\frac{2\,{\mathfrak m} \cos\theta}{r^3}\,{\bf \hat r}+
\frac{{\mathfrak m} \sin\theta}{r^3}\,\boldsymbol{\hat\theta}
\end{equation}
Assuming a spherically-symmetrical magnetic permeability, we have
\begin{equation}{\bf B}=\mu(r) {\bf H} \end{equation}
In analogy with (12), the magnetic field energy integrated over solid
angle is given by
\begin{equation}\frac1{8\pi}
\int_0^{2\pi}\int_0^{\pi}\,{\bf B\cdot H}\,\sin\theta\,d\theta\,d\phi=
\frac{2 \,{\mathfrak m}^2\,\mu(r)}{ r^6 }
\end{equation}
The current density of polarized charge can be found from
\begin{equation}
{\bf J}=\frac{c}{4\pi}\,\nabla\times{\bf B}\approx\frac{\mu'(r)}{r^3}
\end{equation}
An analogous assumption that the magnetic energy density is proportional to the
polarization current density thus implies a magnetic susceptibility of the form
\begin{equation} \mu(r)=\exp(-b^2/r^2)
\end{equation}

Since $\epsilon(r)\to\infty$ and $\mu(r)\to 0$ as $r\to 0$, both $\bf E\cdot D$ and
$\bf B\cdot H$ vanish at the origin. Thus we need not consider contact contributions of
the form $\boldsymbol{\mathfrak p}\,\delta({\bf r})$ or $\boldsymbol{\mathfrak
m}\,\delta({\bf r})$.

\section{Classical Model for the Electron}

Let the electron be pictured as a structureless point charge $e$ with a  magnetic
dipole moment
${\mathfrak m}=e\hbar/2 m c$.  
If the energy is entirely electromagnetic, according to Lorentz-Abraham, the intrinsic
angular momentum should likewise be electromagnetic. In this way we
can sidestep any need to explain how a point particle can have a  spin angular
momentum. (Alternatively, this might be attributed to the motion of the
polarization charge surrounding the electron.) The angular momentum of an
electromagnetic field is given by
\begin{equation}
{\bf S}=\frac{1}{4\pi c}\int\,{\bf r\times (E\times H)}\,d^3{\bf r}
\end{equation}
Identifying this with the electron's spin of one-half, we can write
\begin{equation}
S_z=\frac{1}{4\pi c}\int\, r\sin\theta\,{\bf(E\times H)}_\phi\,d^3{\bf r}=\frac\hbar2
\end{equation}
Using the fields
\begin{equation}
{\bf D}=\frac e{r^2}{\bf \hat r},\; {\bf E}=\frac{\bf D}{\epsilon(r)},\quad
{\bf H}=\frac{2\,{\mathfrak m} \cos\theta}{r^3}\,{\bf \hat r}+
\frac{{\mathfrak m} \sin\theta}{r^3}\,\boldsymbol{\hat\theta}
\end{equation}
with
\begin{equation}{\mathfrak m}=\frac{e\hbar}{2mc}\end{equation}
and the permittivity parametrized as
\begin{equation}\epsilon(r)=e^{a/r}\end{equation}
Eq (21) is satisfied with
\begin{equation}a=\frac23\frac{e^2}{m c^2}=\frac23 r_0 \end{equation}
The electric-field energy then works out to 
\begin{equation}W{\scriptscriptstyle\rm elec}={1\over{8\pi}}\,\int\,{\bf E\cdot D} 
\;d^3{\bf r}=\frac34\,mc^2
\end{equation}
The magnetic contribution must then supply the
remaining quarter of the rest energy:
\begin{equation}W{\scriptscriptstyle\rm mag}={1\over{8\pi}}\,\int\,{\bf B\cdot H} 
\;d^3{\bf r}=\frac14\,mc^2
\end{equation}
With the parametrization $\mu(r)=\exp(-b^2/r^2)$, Eq (27) is satisfied with
\begin{equation}b=\left(\frac{{\mathfrak m}^2\sqrt\pi}{m c^2}\right)^{1/3}=
\frac{\pi^{1/6}}{2^{2/3}\alpha^{2/3}}\,r_0
\end{equation}
where $\alpha=e^2/\hbar\, c$, the fine structure constant.

\section{Conclusion}

We have shown how to accommodate point structures with finite self-energies into
classical electrodynamics  {\it without} altering the equations of Maxwell's theory. 
This is in contrast to earlier attempts of Born[5], Bopp[6] and others,
which involved nonlinear reformulations of the fundamental equations. The key to our
approach is the introduction of constitutive parameters for the electromagnetic
vacuum, which actually has a physical rationale according to quantum field theory. In any
event, our results reduce smoothly to conventional electrodynamics for scales large
compared to
$10^{-13}$ cm. In particular, $\epsilon(r)$ and $\mu(r)$ both rapidly approach their
vacuum values of 1.

The Lorentz-Abraham conjecture, that the electron's
rest mass of 0.511 MeV$/c^2$ is entirely
electromagnetic, is made more plausible by the model we have described.  This is 
consistent as well with the (nearly, if not exactly) zero rest mass of the electron's
uncharged weak isodoublet partner---the neutrino---which can be regarded as an electron
with zero charge. We note also that the neutron-proton mass difference (1.29 MeV/$c^2$)
is of comparable order of magnitude.  The parameters  which we have fit to the
electron's mass, angular momentum and magnetic moment imply a $g$-factor of 2, 
consistent with Dirac's relativistic theory. (We have resisted the temptation to adjust this
to 2.0023, to account for QED radiative corrections.)

Of course,  the {\it   real}  physical electron must ultimately be described by quantum
mechanics or quantum field theory.  Still, a fully consistent classical model can provide a
useful starting point[7].  And classical results do (usually) represent $\hbar\to 0$
limits in quantum theory.  Since it is by no means settled that the current formalism of
quantum electrodynamics is the final theory of the electron, it is worthwhile to explore the
classical limit that some successor theory might also exhibit. Although the infinities
associated with transverse radiation fields do remain,  we have succeeded in eliminating
those of classical origin for a point charge.

\bigskip

\leftline{\bf REFERENCES}

[1] A definitive review of classical electron theories is given by  F.
Rohrlich, {\it        Classical Charged  Particles} (Addison-Wesley, Reading, MA,
1990).

[2] W. Furry and J. R. Oppenheimer, Phys. Rev. {\bf 45}, 245 (1934).

[3] V. F. Weisskopf, Det. Kgl. Danske Videnskab. Selskab. Mat.-Fys.
Medd.  {\bf 14}, 1  (1936). Reprinted in J. Schwinger, {\it        Quantum
Electrodynamics}  (Dover, New York, 1958).

[4] S. M. Blinder, Repts. Math. Phys. {\bf 47}, 269 (2001). Online
version:
\url{http://arXiv.org/find/physics/1/au:+Blinder/0/1/0/past/0/1}

[5] M. Born, Proc. Roy. Soc. {\bf A143}, 410 (1934); M. Born and L.
Infeld,  Proc. Roy. Soc. {\bf A144}, 425 (1934).

[6] F. Bopp, Ann. Phys. {\bf 38}, 345 (1940).

[7] The connection between classical and quantum theories of the
electron is discussed by P. Pearle, ``Classical electron models," in {\it       
Electromagnetism: Paths to Research}, ed D. Teplitz  (Plenum, New York, 1982)
pp. 211-295.

\end{document}